\documentclass[a4paper,11pt]{article}

\usepackage[latin1]{inputenc}
\RequirePackage{vruler}
\usepackage{crop} 
\usepackage{amsmath,amssymb,amsfonts,upref,endnotes}
  \usepackage{marginnote}
 \usepackage{soul}
\RequirePackage{graphicx}
\usepackage[figuresright]{rotating}
\usepackage{color}
 \usepackage{makeidx}
\usepackage{bm,upgreek}
\usepackage{psfrag} 
\usepackage[margin=0.9in]{geometry}

\def\tr{\;\textrm{tr}}
\def\d{\textrm{d}}

\author{Riccardo De Pascalis,  I.\ David Abrahams\thanks{\textbf{Author for correspondence:} \texttt{i.d.abrahams@manchester.ac.uk}},  William J.\ Parnell, \\ \footnotesize{School of Mathematics, University of Manchester,} \\\footnotesize{Oxford Road, Manchester, M13 9PL, United Kingdom}}
\title{On nonlinear viscoelastic deformations --  \\ a reappraisal of Fung's quasilinear viscoelastic model}

 \date{}

\begin{document}
\maketitle
\begin{abstract}
This article offers a reappraisal of Fung's method for quasilinear viscoelasticity. It is shown that a number of negative features exhibited in other works, commonly attributed to the Fung approach, are merely a consequence of the way it has been applied. The approach outlined herein 
is shown to yield improved behaviour, and offers a straightforward scheme for solving a wide range of models.
Results from the new model are contrasted with those in the literature for the case of uniaxial elongation of a bar: for an imposed stretch of an incompressible bar, and for an imposed load. In the last case, a numerical solution to a Volterra integral equation is required to obtain the results. This is achieved by a high order discretisation scheme. Finally, the stretch of a compressible viscoelastic bar is determined for two distinct materials: Horgan-Murphy and Gent.
\end{abstract}

\begin{center}
\textsc{keywords:} viscoelastic, quasilinear, Fung, Pipkin-Roger, Horgan-Murphy, Gent, hyperelastic, biological soft tissue, elastomer, strain energy function
\end{center}

\begin{center}
\textsc{subject}: solid mechanics, material science, engineering
\end{center}

\section{Introduction}

The study of nonlinear viscoelastic deformations of solid materials has a very long history, with a consequent proliferation of a diverse and extensive array of constitutive models. For a comprehensive overview of the topic the reader is directed to the recent paper by Wineman \cite{Wineman09} and related references therein. The subject is still active, with models continuing to be developed across the field, from highly mathematical approaches where implementation is not a concern, to very applied studies where ease of application is essential. Each approach has its advantages: the models which more accurately describe the microphysics tend to prove difficult to employ in practical engineering or biomedical situations, whereas simpler approaches often miss crucial detail. If the viscous model assumes a fading memory effect of the strain history then this usually gives rise mathematically to Volterra-type integral equations, which in the linear case can generally be solved either analytically or numerically.
However, in the nonlinear case (of interest to describe finite deformations) these problems are difficult to solve by any methods, even finite element approaches. It is therefore crucial to develop constitutive models that are simple enough to be amenable to straightforward (and rapid) solution methods, yet include enough detail to capture the important physics underlying these relevant materials.  One such `compromise' approach was offered by Fung \cite{fung81}, in order to study the uniaxial elongation of biological soft tissue. His constitutive assumption, often called \textit{quasilinear viscoelasticity} (QLV) or \textit{Fung's model} of viscoelasticity, assumes that the viscous relaxation rate is independent of the instantaneous local strain.  This model, which is a special case of a more general Pipkin-Rogers constitutive model \cite{pipkin_rogers1968}, predicts at any time a stress that is equal to the instantaneous elastic stress response decreased by an amount depending on the past history, assuming that 
a Boltzmann superposition principle holds.

Fung's quasilinear viscoelastic (QLV) model, which is perhaps the most widely used today, has met with some criticism, mainly because it has been suggested that it does not yield physically `reasonable' behaviour. This apparent deficiency will be discussed below, but the purpose of this article is to reappraise Fung's approach, suggesting how it can be reformulated against existing interpretations in the literature. It will be seen that the form offered herein has similarities to other viscoelastic formulations, including that by Simo \cite{simo1987}, and most importantly, exhibits behaviour that makes it both (relatively) easy to use and physically useful.

We commence with the general tensorial form of (QLV) proposed by Fung in \cite{fung81}:
\begin{equation}\label{quasi_linear_constitutive_law}
\bm{\Uppi}(\mathbf{X},t)=\int_{-\infty}^t \mathbf{G}(t-\tau)\dfrac{\partial \bm{\Uppi}^\textrm{e}(\mathbf{X},\tau)}{\partial  \tau}\ \d \tau,
\end{equation}
where $\mathbf{G}(t)$ is, in comparison with linear theory, the stress relaxation second-order tensor, $\bm{\Uppi}$ denotes the second Piola-Kirchhoff stress, while $\bm{\Uppi}^\textrm{e}$ is an instantaneous strain measure. The latter may be thought of as an equivalent (instantaneous) \textit{elastic} stress, hence the superscript \lq$\textrm{e}$\rq. This tensorial integral identity is the natural generalisation of the simple one-dimensional (1D) relationship proposed by Fung \cite{fung81}, which preserves \textit{objectivity}.

As mentioned, thanks to the relative (apparent) simplicity of Fung's approach over more general nonlinear viscoelastic models, it has proved extremely popular. This is especially true in the biomechanics field where it has been employed to predict the large deformations of soft tissues. However, despite the widespread use of QLV, published studies (especially in the incompressible limit) appear to interpret \eqref{quasi_linear_constitutive_law} in different ways, which then lead to quite different results. In the paragraphs below a number of common approaches are discussed, where the interpretations of the model may be questioned. The main aim of this article is then to re-derive QLV, starting from basic principles, and also to ensure consistency in the limit of infinitesimal deformations, thus recovering the Boltzmann theory of linear viscoelasticity.

The simplest interpretation of Fung's relation is to assume a purely 1D homogenous deformation. This is attractive for its ease of use and has been employed extensively in biomedical applications \cite{abramowitch_woo_clineff_al2004, abramowitch_woo2004, yoo_kim_gupta_demer2009}, for example to model the shearing deformation of brain tissue \cite{rashid_destrade_gilchrist2013}. However, in practical applications, especially for incompressible or near incompressible materials, purely 1D deformations, such as pure uniaxial extension, will not be realisable. Thus, a tensorial form of QLV must be employed. As mentioned, this relation must, at a minimum, satisfy objectivity, and hence the form chosen in \eqref{quasi_linear_constitutive_law}. Nevertheless, a number of authors have erroneously expressed Fung's relation not for the 2nd Piola-Kirchhoff stress, which guarantees objectivity, but in terms of another stress measure. For example, see Fung's original discussion on the subject in \cite{fung81} (page 253), which works with the Kirchhoff stress! Note also that early versions of the viscoelastic model employed in ABAQUS failed to satisfy objectivity, but a recent version [v.6.12] does.

Another approach to quasilinear viscoelasticity sometimes employed in the literature (e.g.\ \cite{miller_chinzei2002, rashid_destrade_gilchrist2012, miller_chinzei1997}) is, in an analogous fashion to nonlinear elasticity, to write the stress in terms of the instantaneous derivative (with respect to the principal stretch) of a strain energy function. Clearly, in this case the strain energy function must be dissipative, and so these authors
express it as a fading-memory integral with integrand given as a hyperelastic strain energy function. This approach may be somewhat hard to justify, and does not allow the user to consider the problem in terms of an auxiliary instantaneous measure of strain, i.e.\ the effective elastic stress term $\bm{\Uppi}^\textrm{e}$ given above.

Perhaps a more subtle point to those mentioned above is the choice of $\bm{\Uppi}^\textrm{e}$ in \eqref{quasi_linear_constitutive_law}. In this article, the reasonable assertion is made that $\bm{\Uppi}^\textrm{e}$ must be zero whenever the deformation is zero. However, this point is often not recognised, especially for incompressible materials, as can be seen, for example, by inspection of the integrands (setting the stretch equal to unity) in the articles by \cite{ciambella2009abaqus} (see equations (8) and (12) therein), \cite{johnson_livesay_woo_Rajagopal1996} (equations (9) and (13)) and \cite {woo_savio_abramowitch_al2006} (equation (9)). Note that this requirement is satisfied in the 1D models in \cite{abramowitch_woo_clineff_al2004}  and \cite{abramowitch_woo2004} but not in \cite{yoo_kim_gupta_demer2009}. If $\bm{\Uppi}^\textrm{e}$ does not vanish for zero deformation then, in general, it can be expected that the actual stress field $\bm{\Uppi}$ will be non-zero prior to the application of the imposed load or stretch, i.e.\ the solution will be non causal! However, for incompressible materials, an arbitrary pressure term (Lagrange multiplier) can be adjusted so that the stress field is indeed causal, but then it must take a specific form at later times. That is, whenever the material is deformed, and then returns back to zero, the stress $\bm{\Uppi}$ will instantaneously return to zero too. This is clearly an unrealistic consequence of the particular choice of model, for, as Fung states (\cite{fung81}, page 228): \textit{`the tensile stress at any time $t$ is equal to the instantaneous [elastic] stress response \ldots decreased by an amount depending on the past history'}. This issue is examined further in \S \ref{elongation}\ref{simple_extension}.

There are other variants of QLV employed in the literature that offer slightly modified forms of the governing equation to that suggested by Fung. The reader is referred, for example, to articles \cite{nekouzadeh_al2007, provenzano_lakes_corr_vanderby2002, pena_al2007}. In this paper, the method employed follows closely that proposed by Fung, but does not exhibit any of the limitations just discussed.

The paper is organised as follows. In the following section, the usual (Boltzmann) linear viscoelastic model is reviewed by way of introduction to a new interpretation of Fung's quasilinear viscoelastic theory, presented in \S \ref{QLV_section}. In \S \ref{elongation}, results from the new model are contrasted with those in the literature for the case of uniaxial elongation of a bar; in \S \ref{elongation}\ref{simple_extension} for an imposed stretch of an incompressible bar, in \S \ref{elongation}\ref{simple_tensile_load} for an imposed load, and in \S \ref{elongation}\ref{simple_extension_compressible} for stretch of a compressible bar. A numerical solution to a Volterra integral equation is required for the results in \S \ref{elongation}\ref{simple_tensile_load}, and the discretisation (time-stepping) procedure used is described in the Appendix \ref{numeric}. Finally, concluding remarks are offered in \S\ref{conclusion}.

\section{Boltzmann's linear viscoelastic law}\label{linear}
It is helpful to commence analysis by recapping the theory of linear viscoelasticity. Under the assumption of isotropy, infinitesimal elastic deformations can be described by the constitutive law
\begin{equation}\label{linear_elasticity}
\bm{\upsigma}(t)=2\mu\left(\bm{\upepsilon}(t)-\dfrac{1}{3}\textrm{tr}\left(\bm{\upepsilon}(t)\right)\mathbf{I}\right)  + \kappa \textrm{tr}\left(\bm{\upepsilon}(t)\right) \mathbf{I}
\end{equation}
where $\bm{\upsigma}$ and $\bm{\upepsilon}$ are the second order stress and strain tensors, respectively, and $\mu$ and $\kappa$ represent the infinitesimal shear modulus and
modulus of compression (or bulk modulus) respectively.  To incorporate viscoelastic behaviour, the most natural extension of \eqref{linear_elasticity} is to assume that the stress \textit{remembers} the past history of the rate of strain with some \textit{fading memory}, and then apply the \textit{superposition principle} (or Boltzmann's principle); hence
\begin{equation}\label{linear_viscoelasticity}
\bm{\upsigma}(t)=2\int_{-\infty}^{t}  \mu(t-s) \dfrac{\partial}{\partial s}\left(\bm{\upepsilon}(s)-\dfrac{1}{3}\left(\textrm{tr}\bm{\upepsilon}(s)\right)\mathbf{I}\right)\ \d s  + \int_{-\infty}^{t} \kappa(t-s)  \dfrac{\partial}{\partial s} \left(\textrm{tr}\bm{\upepsilon}(s)\right) \mathbf{I}\ \d s,
\end{equation}
where $\mu(t), \kappa(t)$ are now time dependent \textit{relaxation} functions, and the lower limit of the integral must be taken from $-\infty$ in order to correctly consider the initial deformation. Integrating \eqref{linear_viscoelasticity} by parts, and assuming that the deformation commences at $t=0$, yields
\begin{multline}\label{linear_viscoelasticity_integrated}
\bm{\upsigma}(t)= 2\mu(0)\left(\bm{\upepsilon}(t)-\dfrac{1}{3}\left(\textrm{tr}\bm{\upepsilon}(t)\right)\mathbf{I}\right)+2\int_{0}^{t} \mu'(t-s)  \left(\bm{\upepsilon}(s)-\dfrac{1}{3}\left(\textrm{tr}\bm{\upepsilon}(s)\right)\mathbf{I}\right)\ \d s  + \\ \kappa(0) \textrm{tr}\bm{\upepsilon}(t)\mathbf{I}+  \int_{0}^{t}    \kappa'(t-s)  \textrm{tr}\bm{\upepsilon}(s)  \mathbf{I}\ \d s,
\end{multline}
where the $'$ denotes differentiation with respect to the argument of the function. Note that this expression incorporates any jump discontinuity when the motion starts. Moreover, the first term in \eqref{linear_viscoelasticity} (or equivalently the first two terms in \eqref{linear_viscoelasticity_integrated}) is trace free, or deviatoric, and accounts for shear deformations and loss, while the second term accounts for the hydrostatic part representing compressive deformations and loss.

Now, in the elastic case \eqref{linear_elasticity}, incompressibility may be considered as the dual limit of $\kappa/\mu\rightarrow\infty$ and $\textrm{tr}\bm{\upepsilon}\rightarrow0$,  which results in the stress-strain relationship
\begin{equation}\label{linear_elasticity_incom}
\bm{\upsigma}(t)=2\mu\bm{\upepsilon}(t)- p(t)\mathbf{I},
\end{equation}
i.e.\ the second term in \eqref{linear_elasticity} yields a finite non-zero limit, where $p(t)$ may be considered as a Lagrange multiplier of incompressibility. For the viscoelastic counterpart, the limit of incompressibility can be  considered in a similar fashion, noting first that $\mu(\infty)$ and $\kappa(\infty)$ are the long time shear and bulk moduli (the equivalent of $\mu,\kappa$ in the elastic case), respectively. Thus, taking $\kappa(\infty)\rightarrow\infty$ (which implies $\kappa(t)\rightarrow\infty$ for all $t$ due to the fading memory assumption) and $\textrm{tr}\bm{\upepsilon}\rightarrow0$ in \eqref{linear_viscoelasticity} (or in \eqref{linear_viscoelasticity_integrated}), it is found that
\begin{equation}
\kappa(0) \textrm{tr}\bm{\upepsilon}(t) +  \int_{0}^{t}    \kappa'(t-s)  \textrm{tr}\bm{\upepsilon}(s)  \ \d s\rightarrow -p(t),
\end{equation}
where the
limit is again assumed to have a non zero finite value, $-p(t)$. Thus, in the limit of incompressibility,  equations \eqref{linear_viscoelasticity} and \eqref{linear_viscoelasticity_integrated} become
\begin{equation}  \label{linear_inc_viscoelasticity}
\bm{\upsigma}(t)=-p(t)\mathbf{I}+2\int_{-\infty}^{t}  \mu(t-s) \dfrac{\partial}{\partial s} \bm{\upepsilon}(s)  \ \d s,
\end{equation}
and
\begin{equation}  \label{linear_inc_viscoelasticity_integrated}
\bm{\upsigma}(t)=-p(t)\mathbf{I}+2\mu(0)\bm{\upepsilon}(t)+2\int_{0}^{t}    \mu'(t-s) \bm{\upepsilon} (s)\ \d s,
\end{equation}
respectively.

\section{Quasilinear Viscoelasticity (QLV)}\label{QLV_section}

When the strain is not infinitesimal, linear theory becomes inappropriate to describe deformations; hence a nonlinear constitutive law has to be considered. As discussed in the introduction, Fung's hypothesis \eqref{quasi_linear_constitutive_law} is examined here as a means of describing the motion of viscous nonlinearly-elastic materials. In his renowned work \cite{fung81}, Fung introduced this quasilinear constitutive model in order to capture the nonlinear stress-strain relationship of living tissues; however, it also has applicability to elastomeric materials.

Before deriving the quasilinear theory, it is useful to introduce some standard definitions and notations.  The deformation gradient tensor $\mathbf{F}$ is defined by
\begin{equation}  \label{deformation_gradient_s}
\mathbf{F}(s)=\Bigg\{\begin{array}{cc}
\mathbf{I}, & \quad s\ \in \  (-\infty ,0), \\
\dfrac{\partial \mathbf{x}(s)}{\partial \mathbf{X}}, &  s\ \in \ [0,t],
\end{array}
\end{equation}
with $\mathbf{x}(s)$ denoting the position of a generic particle $P$ at time $s\ \in \ [0 , t]$, and $\mathbf{X}$  its position at the initial reference time. Note that the start time of the deformation, and any imposed tractions, will be taken as $t=0$. The quantity $J=\textrm{det} \mathbf{F}$, expressing the local volume change, is a constant $J=1$ when the deformation is isochoric. Further, from the deformation gradient tensor $\mathbf{F}$ the left Cauchy-Green tensor $\mathbf{B}=\mathbf{F}\mathbf{F}^{T}$ is obtained, together with its principal invariants
\begin{equation}
I_1=\tr \mathbf{B},\quad I_2=\tfrac{1}{2}[(\tr \mathbf{B})^2-\tr \mathbf{B}^2]=\left(\det\mathbf{B}\right)\tr\left(\mathbf{B}^{-1}\right),\quad I_3=\det \mathbf{B}= J^2,
\end{equation}
which, alternatively, can be expressed in terms of the principal stretches through
\begin{equation}
I_1=\lambda_1^2+\lambda_2^2+\lambda_3^2,\quad
I_2=\lambda_1^2\lambda_2^2 +\lambda_2^2\lambda_3^2+\lambda_1^2\lambda_3^2,\quad  \label{invariants}
I_3=\lambda_1^2 \lambda_2^2 \lambda_3^2.
\end{equation}


Now, Fung \cite{fung81} makes the assumption that the quasilinear viscoelastic stress depends linearly on the (superposed) time history of a related nonlinear elastic response (a nonlinear instantaneous measure of strain). This allows, for example, for incorporation of a finite hyperelastic theory in the analysis.
In index notation, Fung's theory \eqref{quasi_linear_constitutive_law} can be written as\footnote{Note, in order to clarify exposition, the space variable in \eqref{quasi_linear_constitutive_law} is omitted; the same will be done henceforth, as long as no confusion occurs.}
\begin{equation}\label{quasi_linear_constitutive_law_index}
 \Uppi_{ij}(t)=\int_{-\infty}^t G_{ijkl}(t-\tau)\dfrac{\partial  \Uppi_{kl}^\textrm{e}(\tau)}{\partial  \tau}\ \d \tau.
\end{equation}
Fung refers to $G_{ijkl}$ as a \textit{reduced relaxation function tensor}. 
The `crucial' simplification of Fung's theory is that this term is \textit{independent} of the strain. Moreover, if the material is isotropic then $\mathbf{G}$, a tensor of rank four, has just two independent components, being therefore consistent with linear theory\footnote{This is not always taken in account, for example in \cite{drapaca2007nonlinear} the authors introduce an isotropic model which contains only one scalar relaxation function $G$.}.

Following the analysis of the previous section, it is convenient to split the equivalent (instantaneous) \textbf{Cauchy stress} into two parts, one which accounts for \textit{microscopic isochoric deformations} of the material and one that measures purely \textit{compressive deformations}. These two components can be expected to have different instantaneous elastic behaviours as well as distinct relaxation rates, associated for example with the unwinding/unravelling of polymeric fibres as opposed to their stretching. It is assumed that this decomposition can be achieved by taking the deviatoric and hydrostatic components of the equivalent elastic Cauchy stress:
\begin{equation}\label{cauchy_elast_stress}
\mathbf{T}^\textrm{e}=\mathbf{T}^\textrm{e}_\textrm{D}+\mathbf{T}^\textrm{e}_\textrm{H},
\end{equation}
which can be expressed as
 \begin{equation}\label{cauchy_elast_stress_split}
\mathbf{T}^\textrm{e}_\textrm{H}=\dfrac{1}{3}\tr\left(\mathbf{T}^\textrm{e}\right) \mathbf{I},\quad \mathbf{T}^\textrm{e}_\textrm{D}=\mathbf{T}^\textrm{e}-\dfrac{1}{3}\tr\left( \mathbf{T}^\textrm{e}\right) \mathbf{I}.
\end{equation}
The split of equation \eqref{cauchy_elast_stress} into shear and dilatational components is the hyperelastic analogue of the linear stress-strain law \eqref{linear_elasticity}. However, it cannot be generalised to viscoelasticity, as in \eqref{linear_viscoelasticity}, because it would not preserve \textit{objectivity}. Instead, the second Piola-Kirchhoff stress tensor associated with \eqref{cauchy_elast_stress} has to be introduced, which is defined by
\begin{equation}\label{piolaK_elastic_stress_general}
\bm{\Uppi}^\textrm{e}= J \mathbf{F}^{-1} \mathbf{T}^\textrm{e} \mathbf{F} ^{-T},
\end{equation}
with
\begin{equation}\label{piola_elast_stress_h+d}
\bm{\Uppi}^\textrm{e}= \bm{\Uppi}_\textrm{D}^\textrm{e}+ \bm{\Uppi}_\textrm{H}^\textrm{e},
\end{equation}
in which
\begin{equation}\label{piolaK_h+d}
\bm{\Uppi}_\textrm{D}^\textrm{e}= J \mathbf{F}^{-1} \mathbf{T}_\textrm{D}^\textrm{e} \mathbf{F} ^{-T},\quad
\bm{\Uppi}_{\textrm{H}}^\textrm{e}= J \mathbf{F}^{-1} \mathbf{T}_\textrm{H}^\textrm{e} \mathbf{F} ^{-T}.
\end{equation}
It must be emphasised that the subscripts $\textrm{D}$ and $\textrm{H}$ \textbf{do not} refer to the deviatoric and hydrostatic parts of the second Piola-Kirchhoff stress, but correspond to the second Piola-Kirchhoff stress of the deviatoric and hydrostatic Cauchy stress components, respectively.  Assuming a superposition principle as for the linear case, it is now possible to introduce an objective viscoelastic law, relating the second Piola-Kirchhoff stress to the past history of the nonlinear rate of strain measure. This is taken as
\begin{equation}\label{piolaK_viscoelastic_stress}
\bm{\Uppi}(t)= \int_{-\infty}^t   \mathcal{D}(t-s) \dfrac{\partial}{\partial s} 
\bm{\Uppi}^\textrm{e}_\textrm{D}(s)\ \d s    +  \int_{-\infty}^t    \mathcal{H}(t-s) \dfrac{\partial}{\partial s} 
\bm{\Uppi}^\textrm{e}_\textrm{H}(s)\ \d s,
\end{equation}
where now $\mathcal{D}(t),\mathcal{H}(t)$ are two scalar (independent) reduced relaxation functions (with $\mathcal{D}(0)=\mathcal{H}(0)=1$ without loss of generality). The latter relaxation functions relate to the inherent viscous processes involved with instantaneous shear and compressional (volumetric) deformations, respectively. Clearly, if the material was anisotropic then a more complex tensorial relaxation function would be required. Further, pre-multiplying by $J^{-1} \mathbf{F}$ and post-multiplying by $\mathbf{F}^T$ yields the Cauchy viscoelastic stress
\begin{multline}\label{cauchy_viscoelastic_stress}
\mathbf{T}(t)= J^{-1}\mathbf{F}(t) \left(\int_{-\infty}^t  \mathcal{D}(t-s) \dfrac{\partial}{\partial s} 
\bm{\Uppi}^\textrm{e}_\textrm{D}(s)\ \d s \right) \mathbf{F}^T(t)  +  \\ J^{-1}\mathbf{F}(t) \left(\int_{-\infty}^t  \mathcal{H}(t-s)
\dfrac{\partial}{\partial s} 
\bm{\Uppi}^\textrm{e}_\textrm{H}(s)\ \d s\right) \mathbf{F}^T(t),
\end{multline}
and integrating by parts, following \eqref{linear_inc_viscoelasticity_integrated}, gives
\begin{multline}\label{cauchy_viscoelastic_stress_integrated}
\mathbf{T}(t)= J^{-1}\mathbf{F}(t) \left(\bm{\Uppi}^\textrm{e}_\textrm{D}(t)+ \int_{0}^t  \mathcal{D}'(t-s)\bm{\Uppi}^\textrm{e}_\textrm{D}(s)\ \d s \right) \mathbf{F}^T(t)  +  \\ J^{-1}\mathbf{F}(t) \left( \bm{\Uppi}^\textrm{e}_\textrm{H}(t)+ \int_{0}^t  \mathcal{H}'(t-s)  \bm{\Uppi}^\textrm{e}_\textrm{H}(s)\ \d s\right) \mathbf{F}^T(t).
\end{multline}

As mentioned earlier, to allow for large (nonlinear) deformations, a hyperelastic theory can be employed assuming that the measure of the \textit{effective elastic stress} $\bm{\Uppi}^\textrm{e}$ is derived from an elastic potential. Let us then specialise equations \eqref{cauchy_elast_stress}-\eqref{cauchy_viscoelastic_stress_integrated}, considering the existence of a strain energy function (SEF) $W$, say, which in the isotropic case is dependent on the principal invariants of the deformation $I_1,I_2,I_3$, or of the principal stretches $\lambda_1,\lambda_2,\lambda_3$:
\begin{equation}
W=W(I_1,I_2,I_3)=\tilde{W}(\lambda_1,\lambda_2,\lambda_3).
\end{equation}
The general form of elastic Cauchy stress may be written (see e.g.\ \cite{truesdell-noll92, depascalis2010}) as
\begin{equation} \label{T_Cayley-Hamilton}
\mathbf{T}^\textrm{e}=\beta_0 \mathbf{I}+\beta_1\mathbf{B}+\beta_{-1} \mathbf{B}^{-1},
\end{equation}
where $\beta_j=\beta_j(I_1,I_2,I_3)$  are so-called  material (or elastic) response functions. In terms of the strain energy function they are given by
\begin{align}
&\beta_0(I_1,I_2, I_3)=\frac{2}{J} \left[I_2  W_2 +I_3  W_3 \right], \nonumber \\  & \beta_{1}(I_1,I_2, I_3)= \frac{2}{J}   W_1, \label{elastic_response_functions}  \\
&\beta_{-1}(I_1,I_2, I_3)= -2 J  W_2, \nonumber
\end{align}
where
\begin{equation}\label{W_derivs}
W_k= \dfrac{\partial W}{\partial I_k},\quad k=1,2,3.
\end{equation}
It is straightforward to calculate the trace of $\mathbf{T}^\textrm{e}$,
\begin{equation}
\tr\mathbf{T}^\textrm{e}=3\beta_0+I_1 \beta_1+ I_2/I_3 \beta_{-1}
\end{equation}
and so the deviatoric and hydrostatic elastic Cauchy stress components in \eqref{cauchy_elast_stress_split} become
\begin{align}\label{h+d}
&\mathbf{T}_\textrm{D}^\textrm{e}=\dfrac{2}{ J}\left[\dfrac{1}{3}\left( I_2  W_2-  I _1  W_1\right)\mathbf{I}+ W_1 \mathbf{B}- I_3  W _2 \mathbf{B}^{-1}\right], & \mathbf{T}_\textrm{H}^\textrm{e}=\dfrac{2}{J} \left(\dfrac{2}{3} I_2 W_2+ \dfrac{1}{3} I_1 W_1+ I_3 W_3\right)    \mathbf{I}.
\end{align}
From this the second Piola-Kirchhoff counterparts, \eqref{piolaK_h+d}, are given by
\begin{align}
\bm{\Uppi}_\textrm{D}^\textrm{e}&=2\left[\dfrac{1}{3}\left( I_2  W_2-  I _1  W_1\right) \mathbf{F} ^{-1}\mathbf{F}^{-T}+ W_1 \mathbf{I}- I_3  W _2 \left( \mathbf{F} ^{-1}\mathbf{F}^{-T}\right)^2\right], \label{piolaK_d_SEF}\\  \bm{\Uppi}_{\textrm{H}}^\textrm{e}&=2 \left(\dfrac{2}{3} I_2 W_2+ \dfrac{1}{3}I_1 W_1+ I_3 W_3\right) \mathbf{F} ^{-1}\mathbf{F}^{-T}, \label{piolaK_h_SEF}
\end{align}
from which the viscoelastic stress is obtained via \eqref{piolaK_viscoelastic_stress}, or for the Cauchy stress from \eqref{cauchy_viscoelastic_stress_integrated}.

Note that (as required) objectivity is now preserved for the viscoelastic model \eqref{cauchy_viscoelastic_stress}, but the first part is \textit{not} deviatoric and so the second term is not purely hydrostatic. In fact, in general both the compressive and shear components of the shear history contribute to the deviatoric and hydrostatic parts of $\mathbf{T}$. However, the main point to note is that when there is no deformation, i.e.\ the principal stretches are unity, $\lambda_j=1, j=1,2,3$ and $\mathbf{B}\equiv \mathbf{I}$, then $I_1=I_2=3, I_3=1$ which reveals that the effective deviatoric elastic stress $\mathbf{T}_\textrm{D}^\textrm{e}$ vanishes. Similarly, $\mathbf{T}_\textrm{H}^\textrm{e}$ vanishes as the strain energy function has always to satisfy the additional relation (see e.g.\ \cite{truesdell-noll92})
\begin{equation}
W_1(3,3,1)+ 2 W_2(3,3,1)+W_3(3,3,1)\equiv 0.
\end{equation}
Thus, both terms $\bm{\Uppi}_\textrm{D}^\textrm{e}$ and $\bm{\Uppi}_\textrm{H}^\textrm{e}$ vanish as $\lambda_i\rightarrow 1$, and hence, as there is no applied stress until the initial time $t=0$, justifies taking the integration range for the pair of integrals in \eqref{cauchy_viscoelastic_stress_integrated} as $0$ to $t$.


If it proves convenient to express the strain energy function in term of the principal stretches $W=\tilde{W}(\lambda_1,\lambda_2,\lambda_3)$, then for diagonal $\mathbf{F}$ the quasilinear viscoelastic stress relation \eqref{cauchy_viscoelastic_stress_integrated} may be rewritten as
\begin{multline}\label{cauchy_viscoelastic_stress_integrated_stretches}
T_i(t)= J^{-1}\lambda_i(t)^2 \left(\Uppi^\textrm{e}_{\textrm{D}i}(t)+ \int_{0}^t  \mathcal{D}'(t-s)\Uppi^\textrm{e}_{\textrm{D}i}s)\ \d s \right)    +  \\ J^{-1}\lambda_i(t)^2\left( \Uppi^\textrm{e}_{\textrm{H}i}(t)+ \int_{0}^t  \mathcal{H}'(t-s)   \Uppi^\textrm{e}_{\textrm{H}i}(s)\ \d s\right),
\end{multline}
in which
\begin{equation}
\Pi^\textrm{e}_{\textrm{D}i}=\dfrac{\tilde{W}_i}{\lambda_i}-\dfrac{1}{3\lambda_i^2}\sum_{j=1}^3\lambda_j\tilde{W}_j,\qquad  \Pi^\textrm{e}_{\textrm{H}i}=  \dfrac{1}{3\lambda_i^2}\sum_{j=1}^3\lambda_j\tilde{W}_j,
\end{equation}
where now $\tilde{W}_i$ refers to the derivative $\partial \tilde{W}/\partial\lambda_i$.

A final consideration of this section is the constraint of incompressibility for all possible deformations, i.e.\ $J\equiv1$. In this limit the speed of propagation of compressive disturbances tends to infinity, and hence the relaxation time for viscous dilatational motions can be assumed to tend to zero.  Therefore, the second term in \eqref{cauchy_viscoelastic_stress} (or \eqref{cauchy_viscoelastic_stress_integrated}), in an analogous fashion to that for linear elasticity \eqref{linear_inc_viscoelasticity_integrated}, reduces to
\begin{equation}
J^{-1}\mathbf{F}(t) \left(  \bm{\Uppi}^\textrm{e}_\textrm{H}(t)+ \int_{0}^t  \mathcal{H}'(t-s)  \bm{\Uppi}^\textrm{e}_\textrm{H}(s)\ \d s\right) \mathbf{F}^T(t)\rightarrow -p(t) \mathbf{I},
\end{equation}
where $p(t)$ can be considered as a Lagrange multiplier. Note that this expression in itself is not hydrostatic. It contains both a hydrostatic term, and a component that combines with the first term in \eqref{cauchy_viscoelastic_stress_integrated} to make it deviatoric; hence
  \begin{equation}\label{cauchy_viscoelastic_stress_inc}
\mathbf{T}(t)= \mathbf{F}(t) \left(  \bm{\Uppi}^\textrm{e}_\textrm{D}(t)+ \int_{0}^t  \mathcal{D}'(t-s) \bm{\Uppi}^\textrm{e}_\textrm{D}(s)\ \d s \right) \mathbf{F}^T(t) -p \mathbf{I},
\end{equation}
where now from the first equation in \eqref{piolaK_d_SEF},
\begin{equation} \label{dev_piola_viscoelastic_stress_inc}
\bm{\Uppi}^\textrm{e}_\textrm{D}(t)=2\left[\left(\dfrac{ I_2}{3} W_2-\dfrac{I_1}{3} W_1\right)\mathbf{F} ^{-1}\mathbf{F}^{-T}+  W_1 \mathbf{I}- W_2 \left(\mathbf{F} ^{-1}\mathbf{F}^{-T}\right)^{2}\right].
\end{equation}

\section{Uniaxial elongation}\label{elongation}
One of the simplest, useful and hence most popular experiments to measure the properties of a material is to subject a specimen to a simple elongation test. There is a huge literature of results on such a deformation, and it is therefore useful to examine this homogeneous solution, comparing our result with extant published work. To appreciate the general quasilinear viscoelastic model developed herein, and the approach needed to obtain a solution, three specific cases will be examined. First, in \S \ref{elongation}\ref{simple_extension} a uniaxial stretch is imposed on the specimen, with the stress determined as a relaxing function of the history of the stretch. For comparison, both Yeoh and Mooney-Rivlin incompressible strain energy functions are considered. In \S \ref{elongation}\ref{simple_tensile_load} a tensile load is imposed, and the stretch history (creep) determined from this. In this case the integral equation must be solved numerically, which is the focus of discussion in Appendix \ref{numeric}. Finally, in \S \ref{elongation}\ref{simple_extension_compressible}, the procedure in \S \ref{elongation}\ref{simple_extension} is repeated for a compressible material, with the effect of compressibility on the viscoelastic behaviour highlighted.

\subsection{Simple extension} \label{simple_extension}
In the case of simple extension, assumed homogeneous (in space) and incompressible, the principal stretches may be specified as
\begin{equation}
x_1(t)=\lambda_1(t) X_1,\quad x_2(t)=\lambda_2(t) X_2,\quad x_3(t)=\lambda_2(t) X_3,
\end{equation}
where $(X_1,X_2,X_3)$  and $(x_1,x_2,x_3)$ are the Cartesian coordinates in the undeformed and deformed state respectively, and the stresses are
\begin{equation}\label{simple_load}
T_{11}(t)=T(t), \quad T_{22}(t)=T_{33}(t)=0,\quad T_{ij}=0\ (i\neq j),
\end{equation}
having assumed that the lateral surfaces are free of stress. The deformation gradient $F_{ij}=\partial x_i/\partial x_j$ is of diagonal form:
\begin{equation}\label{F}
\mathbf{F}(t)=\textrm{diag}\left(\lambda_1(t),\lambda_2(t),\lambda_2(t)\right)
\end{equation}
and the constraint of incompressibility, $J=1$, gives a relationship between the stretches $\lambda_1$ and $\lambda_2$, in particular $\lambda_2=\lambda_1^{-1/2}$. Setting $\lambda_1=\lambda$, the deformation gradient tensor $\mathbf{F}$ and the left Cauchy Green tensor $\mathbf{B}= \mathbf{F} \mathbf{F}^T$ become
\begin{equation}\label{F_B}
\mathbf{F}(t)=\textrm{diag}\left(\lambda(t),\lambda^{-1/2}(t),\lambda^{-1/2}(t) \right), \quad \mathbf{B}(t)=\textrm{diag}\left(\lambda^2(t),\lambda^{-1}(t),\lambda(t)^{-1} \right).
\end{equation}
The principal invariants are therefore
\begin{align}
& I_1=\lambda^2+\dfrac{2}{\lambda},\qquad I_2=2\lambda+\dfrac{1}{\lambda^2},\qquad I_3=1, \label{invariants_simple_extens}
\end{align}
and hence taking the diagonal terms of equation \eqref{cauchy_viscoelastic_stress_inc}, yields the principal Cauchy stresses,
\begin{align}\
T_{11}(t)=T(t)&= \lambda^2(t) \left( \Pi ^\textrm{e}_{\textrm{D}11}(t)+ \int_{0}^t  \mathcal{D}'(t-s)  \Pi^\textrm{e}_{\textrm{D}11}(s)\ \d s \right)  -p, \label{stress_components_simple_1}   \\
T_{22}(t)=T_{33}(t)=0&= \lambda^{-1}(t) \left( \Pi ^\textrm{e}_{\textrm{D}22}(t)+ \int_{0}^t  \mathcal{D}'(t-s)  \Pi^\textrm{e}_{\textrm{D}22}(s)\ \d s \right)  -p,  \label{stress_components_simple_2}
\end{align}
respectively. Thus, by subtraction, the lagrange multiplier $p$ can be eliminated, and so it is possible to rewrite the stress-strain relationship \eqref{stress_components_simple_1}--\eqref{stress_components_simple_2} as
\begin{multline}\label{stress_strain_qlv}
T(t) = \lambda^2(t) \Pi ^\textrm{e}_{\textrm{D}11}(t)-\dfrac{1}{\lambda(t)} \Pi ^\textrm{e}_{\textrm{D}22}(t) +
\int_{0}^t  \mathcal{D}'(t-s)  \left(\lambda^2(t) \Pi ^\textrm{e}_{\textrm{D}11}(s)-\dfrac{1}{\lambda(t)} \Pi ^\textrm{e}_{\textrm{D}22}(s)\right)\ \d s,
\end{multline}
where from \eqref{dev_piola_viscoelastic_stress_inc}
\begin{align} \label{piola_deviatoric_component_11}
\Pi^\textrm{e}_{\textrm{D}11}&=2\left[\dfrac{2}{3}\left(W_1+\dfrac{W_2}{\lambda}\right)\left(1-\dfrac{1}{\lambda^3}\right)\right], \\[3mm]
\label{piola_deviatoric_component_22}
\Pi^\textrm{e}_{\textrm{D}22}&=2\left[\dfrac{1}{3}\left(W_1+\dfrac{W_2}{\lambda}\right)\left(1-  \lambda^3 \right)\right].
\end{align}

Note that these equations \eqref{piola_deviatoric_component_11}, \eqref{piola_deviatoric_component_22} depend on the specific choice of strain energy function $W$,
and so it is useful to examine two specific examples.
Let us start by assuming the instantaneous response is modelled by a two-term Yeoh strain energy function \cite{yeoh1993}
\begin{equation}\label{W_yeoh}
W=\frac\mu4 \left(2 (I_1-3) + \alpha (I_1-3)^2 \right),
\end{equation}
which yields the stress-strain relationship
\begin{equation} \label{yeoh_stress-strain}
\mathbf{T}=-p\mathbf{I}+\mu (1-3\alpha+\alpha  I_1) \mathbf{B},
\end{equation}
where $\alpha$ is a positive constant and $\mu$ is the shear modulus from infinitesimal theory. Thus,
\begin{align}
2W_1=\mu (1-3\alpha+\alpha I_1) , \quad W_2=0.
\end{align}
It is straightforward to obtain from \eqref{stress_strain_qlv} the relation
\begin{multline}\label{stress_strain_qlv_yeoh}
T(t)/\mu=  k(t) \left(\lambda(t)-\dfrac{1}{\lambda^2(t)}\right) + \\ \int_{0}^t  \mathcal{D}'(t-s) k(s) \left(\lambda^2(t) \left[\dfrac{2}{3} \left(\dfrac{1}{\lambda(s)}-\dfrac{1}{\lambda^4(s)}\right)\right]-\dfrac{1}{\lambda(t)}\left[\dfrac{1}{3} \left(\dfrac{1}{\lambda(s)}-\lambda^2(s) \right)\right]\right)\ \d s,
\end{multline}
where
\begin{equation}\label{k_function}
k(t)=2\alpha +\left(1-3\alpha\right) \lambda(t)+\alpha \lambda^3(t).
\end{equation}
A second example is the instantaneous response modelled by a Mooney-Rivlin strain energy function:
\begin{equation}\label{MR_yeoh}
W=\frac\mu2 \left(\frac12+\gamma\right) (I_1-3) + \frac\mu2 \left(\frac12-\gamma\right) (I_2-3),
\end{equation}
which yields the stress-strain relationship
\begin{equation} \label{MR_stress-strain}
\mathbf{T}=-p\mathbf{I}+ \mu\left[\left(\dfrac{1}{2}+\gamma\right) \mathbf{B} - \left(\dfrac{1}{2}-\gamma\right) \mathbf{B}^{-1}\right],
\end{equation}
where $\gamma$ is a constant in the range $-1/2\leq\gamma\leq 1/2$. Note that \eqref{MR_stress-strain} reduces to the neo-Hookean model when $\gamma=1/2$. The partial derivatives of $W$ are
\begin{align}
2W_1= \mu\left(\dfrac{1}{2}+\gamma\right) ,\quad 2W_2= \mu\left(\dfrac{1}{2}-\gamma\right),
\end{align}
and  so \eqref{stress_strain_qlv} becomes, after simplification,
\begin{multline}\label{stress_strain_qlv_MR}
T(t)/\mu= \dfrac{1}{2} \ell(s)  \left(\lambda(t)-\dfrac{1}{\lambda^2(t)}\right) + \\ \dfrac{1}{6}\int_{0}^t    \mathcal{D}'(t-s) \ell(s) \left[2\lambda^2(t) \left(\dfrac{1}{\lambda(s)}-\dfrac{1}{\lambda^4(s)} \right)+\dfrac{1}{\lambda(t)} \left( \lambda^2(s)-\dfrac{1}{\lambda(s)} \right)\right]\ \d s,
\end{multline}
in which
\begin{equation}\label{l_function}
\ell(t)=\left[1-2 \gamma +\lambda(t)(1+2 \gamma )\right].
\end{equation}

\begin{figure}
\psfrag{0}[c][b]{\scriptsize{$0$}}
\psfrag{5}[c][b]{\scriptsize{$5$}}
\psfrag{10}[c][b]{\scriptsize{$10$}}
\psfrag{15}[c][b]{\scriptsize{$15$}}
\psfrag{20}[c][b]{\scriptsize{$20$}}
\psfrag{25}[c][b]{\scriptsize{$25$}}
\psfrag{1.05}[r][c]{\scriptsize{$1.05$}}
\psfrag{1.10}[r][c]{\scriptsize{$1.10$}}
\psfrag{1.15}[r][c]{\scriptsize{$1.15$}}
\psfrag{1.20}[r][c]{\scriptsize{$1.20$}}
\psfrag{1.25}[r][c]{\scriptsize{$1.25$}}
\psfrag{1.30}[r][c]{\scriptsize{$1.30$}}
\psfrag{1.35}[r][c]{\scriptsize{$1.35$}}
\psfrag{-0.1}[r][c]{\scriptsize{$-0.1$}}
\psfrag{0.1}[r][c]{\scriptsize{$0.1$}}
\psfrag{0.2}[r][c]{\scriptsize{$0.2$}}
\psfrag{0.3}[r][c]{\scriptsize{$0.3$}}
\psfrag{0.4}[r][c]{\scriptsize{$0.4$}}
\psfrag{0.5}[r][c]{\scriptsize{$0.5$}}
\psfrag{0.6}[r][c]{\scriptsize{$0.6$}}
\psfrag{0.7}[r][c]{\scriptsize{$0.7$}}
\psfrag{0.8}[r][c]{\scriptsize{$0.8$}}
\psfrag{0.1}[r][c]{\scriptsize{$0.1$}}
\psfrag{gm13}[lb]{\tiny{$\gamma=-1/3$}}
\psfrag{g16}[lb]{\tiny{$\gamma=1/6$}}
\psfrag{al0}[lb]{\tiny{$\alpha=0$}}
\psfrag{al1}[lb]{\tiny{$\alpha=1$}}
\psfrag{al2}[lb]{\tiny{$\alpha=2$}}
\psfrag{aa}[c]{\footnotesize{a}}
\psfrag{bb}[c]{\footnotesize{b}}
\psfrag{cc}[cl]{\footnotesize{c}}
\psfrag{dd}[c]{\footnotesize{d}}
\psfrag{EE}[bl]{\footnotesize{$\lambda$}}
\psfrag{FF}[cb]{\footnotesize{$T/\mu$}}
\psfrag{B}[cb]{\footnotesize{$T/\mu$}}
\psfrag{A}[bl]{\footnotesize{$t$}}
\psfrag{y}[bl]{\footnotesize{$\lambda$}}
\psfrag{x}[bl]{\footnotesize{$t$}}
	\centering
		\includegraphics[scale=0.65]{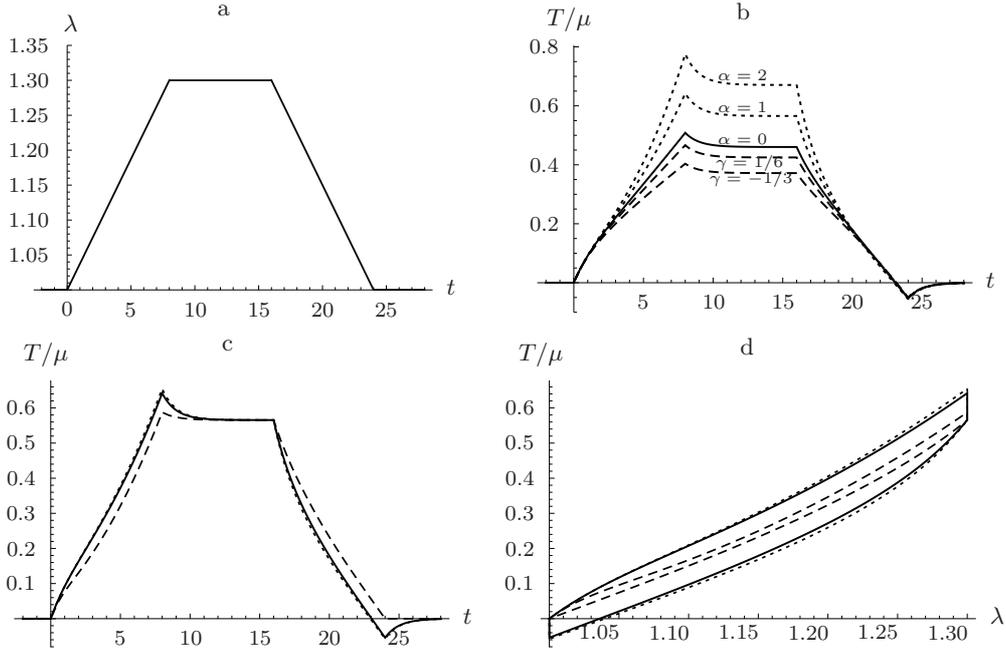}
\caption{The imposed stretch history is shown in graph (a). The resultant dimensionless stress $T/\mu$ is plotted in graph (b), found for the Yeoh model \eqref{stress_strain_qlv_yeoh} (dotted) and for the Mooney-Rivlin material \eqref{stress_strain_qlv_MR} (dashed) where the continuous curve is the neo-Hookean limit $\alpha=0$ (or $\gamma=1/2$). In graph (c), $T/\mu$ is plotted from the predictions of \eqref{abaqus} (dotted), \eqref{QLV_ciambella} (dashed) and \eqref{stress_strain_qlv_yeoh} (continuous), respectively. In graph (d), the dimensionless stress, $T/\mu$, is plotted against stretch, $\lambda$, from the predictions of \eqref{abaqus} (dotted), \eqref{QLV_ciambella} (dashed) and \eqref{stress_strain_qlv_yeoh} (continuous).}
	\label{comparisonstress}
\end{figure}

We can compare these two viscoelastic models by considering an `experiment' where the stretch is imposed and the stress is measured. To specify matters, the stress relaxation function is chosen to be
the classical one-term Prony series
\begin{equation}\label{one_term_prony_series}
\mathcal{D}(t) = \dfrac{\mu_\infty}{\mu}+\left(1-\dfrac{\mu_\infty}{\mu}\right)  e^{-t/\tau},
\end{equation}
where $\mu$, $\mu_\infty$ are the infinitesimal shear modulus and the long-time infinitesimal shear modulus, respectively, and $\tau$ is the relaxation time. These are set as
\begin{equation}\label{constants}
\mu_\infty/\mu=0.5,\quad \tau=1.0s.
\end{equation}
A dynamic imposed stretch history (shown in graph (a) of Fig.\ \ref{comparisonstress}) is applied. The time variation is assumed slow so that inertial terms in the balance equations can be neglected.
Graph (b) in Fig. \ref{comparisonstress} shows the resultant stress predictions $T/\mu$  for the Yeoh hyperelastic model \eqref{stress_strain_qlv_yeoh} (dotted curves with $\alpha=1,2$) and for the Mooney-Rivlin hyperelastic model \eqref{stress_strain_qlv_MR} (dashed curves with $\gamma=1/6,-1/3$). It is interesting to observe that the two hyperelastic models depart from the solid curve, obtained when the strain energy function is of the neo-Hookean type, i.e.\ $\alpha=0,\gamma=1/2$, in different ways. The Yeoh material is found to harden as the parameter $\alpha$ increases, whereas the effect of decreasing $\gamma$ from $1/2$ leads to a softening of the Money-Rivlin material's behaviour.

As mentioned in the introduction, it is useful to contrast the results presented here with those discussed, for example, by Ciambella et al.\ \cite{ciambella2009abaqus}. In that article the authors employed a form of QLV, which, with the incompressible Yeoh SEF, yields the stretch-stress law (see equation (16) in \cite{ciambella2009abaqus}):
\begin{multline}\label{QLV_ciambella}
T(t)/\mu= \left[\lambda(t)-\lambda^{-2}(t)\right] \left[2\alpha+(1-3\alpha) \lambda(t)+\alpha \lambda^3(t)\right]+ \\
 \left[\lambda^2(t)-\lambda^{-1}(t)\right]  \int_0^t \mathcal{D}'(t-s)  \lambda^{-1}(s) \left[2\alpha+(1-3\alpha) \lambda(s)+\alpha \lambda^3(s)\right]\ \d s.
\end{multline}
Similar equations can be derived from the constitutive models in \cite{johnson_livesay_woo_Rajagopal1996} or \cite{woo_savio_abramowitch_al2006}.
In Ciambella et al.'s article, their equation \eqref{QLV_ciambella} was compared with that offered by the formulation used in the ABAQUS finite element analysis (FEA) package (Version 6.7, see \cite{hibbit2007abaqus}), namely
\begin{multline}\label{abaqus}
T(t)/\mu = \left[\lambda(t)-\lambda^{-2}(t)\right] \left[2\alpha+(1-3\alpha) \lambda(t)+\alpha \lambda^3(t)\right]+ \\
\int_0^t \mathcal{D}'(t-s) \left[\lambda(s)-\lambda^{-2}(s)\right] \left[2\alpha+(1-3\alpha) \lambda(s)+\alpha \lambda^3(s)\right]\ \d s.
\end{multline}
The three alternative viscoelastic expressions, \eqref{abaqus}, \eqref{QLV_ciambella} and our result \eqref{stress_strain_qlv_yeoh}, may be conveniently compared by again imposing a stretch and determining the resultant dimensionless stress $T/\mu$. Employing the same stretch history as above (graph (a) in Fig.\ \ref{comparisonstress}), the relaxation function in \eqref{one_term_prony_series} with parameters as in \eqref{constants} and $\alpha=1.0$, yields the curves in graph (c) predicted from \eqref{abaqus} (ABAQUS: dotted), \eqref{QLV_ciambella} (Ciambella et al.: dashed) and \eqref{stress_strain_qlv_yeoh} (De Pascalis et al.: continuous), respectively. An alternative way of representing this deformation is via a stretch-stress diagram, see graph (d) in Fig.\ \ref{comparisonstress}. It is clear that the present QLV approach, illustrated by the continuous curve in both figures, gives a result that lies remarkably close to that found from the ABAQUS model, whereas Ciambella et al.'s solution is quite distinct. The latter model was developed by the authors in order to address the deficiency they highlighted regarding version 6.7 of ABAQUS, namely that the viscoelastic law used was not, in general, objective. However, their result also exhibits a substantial limitation: it predicts instantaneous zero stress whenever $\lambda$ recovers back to unity after some (arbitrary) deformation, thus showing no `memory' of the history of the stress in this situation. This is at variance with physically observed behaviour, and in particular that found for linear viscoelasticity.

\subsection{Simple tensile load} \label{simple_tensile_load}
Rather than an imposed stretch, perhaps a more important experiment for measuring viscoelastic properties of materials is application of a simple tensile load: a slowly varying uniaxial stress is imposed on the body for which the resultant stretch is measured over time. For linear viscoelasticity, equation \eqref{linear_viscoelasticity} can be inverted to yield a straightforward creep relation to determine the strain. However, for nonlinear viscoelasticity, the solution procedure is somewhat more difficult as inversion is not possible. A typical QLV relation is given in \eqref{stress_strain_qlv_MR}, which may be considered in the general form
\begin{equation}\label{general_num_eq}
T(t)=g(\lambda(t))+\sum_{j=1}^{N} f_j(\lambda(t)) \int_0^t \mathcal{D}'(t-s) h_j(\lambda(s))\ \d s.
\end{equation}
Although the integrand is of separable type, the presence of the various $f_j(\lambda(t))$ terms means that an inversion operator cannot be introduced. Instead, a numerical scheme must be employed that can evaluate the stretch as a function of time, subject to a prescribed stress. A suitable numerical discretisation procedure is offered in Appendix \ref{numeric}, which has error $O(\delta t^4)$, where $\delta t$ is the step size, and so gives rapid convergence.

\begin{figure}
\psfrag{0}[c][b]{\scriptsize{$0$}}
\psfrag{5}[c][b]{\scriptsize{$5$}}
\psfrag{10}[c][b]{\scriptsize{$10$}}
\psfrag{15}[c][b]{\scriptsize{$15$}}
\psfrag{20}[c][b]{\scriptsize{$20$}}
\psfrag{25}[c][b]{\scriptsize{$25$}}
\psfrag{30}[c][b]{\scriptsize{$30$}}
\psfrag{1.1}[r][c]{\scriptsize{$1.1$}}
\psfrag{1.2}[r][c]{\scriptsize{$1.2$}}
\psfrag{1.3}[r][c]{\scriptsize{$1.3$}}
\psfrag{1.4}[r][c]{\scriptsize{$1.4$}}
\psfrag{0.1}[r][c]{\scriptsize{$0.1$}}
\psfrag{0.2}[r][c]{\scriptsize{$0.2$}}
\psfrag{0.3}[r][c]{\scriptsize{$0.3$}}
\psfrag{0.4}[r][c]{\scriptsize{$0.4$}}
\psfrag{0.5}[r][c]{\scriptsize{$0.5$}}
\psfrag{0.6}[r][c]{\scriptsize{$0.6$}}
\psfrag{0.7}[r][c]{\scriptsize{$0.7$}}
\psfrag{gm13}[lb]{\tiny{$\gamma=-1/3$}}
\psfrag{g16}[lb]{\tiny{$\gamma=1/6$}}
\psfrag{al0}[lb]{\tiny{$\alpha=0$}}
\psfrag{al1}[lb]{\tiny{$\alpha=1$}}
\psfrag{al2}[lb]{\tiny{$\alpha=2$}}
\psfrag{y}[bc]{\footnotesize{$T/\mu$}}
\psfrag{A}[bl]{\footnotesize{$t$}}
\psfrag{B}[bl]{\footnotesize{$\lambda$}}
\psfrag{x}[bl]{\footnotesize{$t$}}
	\centering
		\includegraphics[scale=0.62]{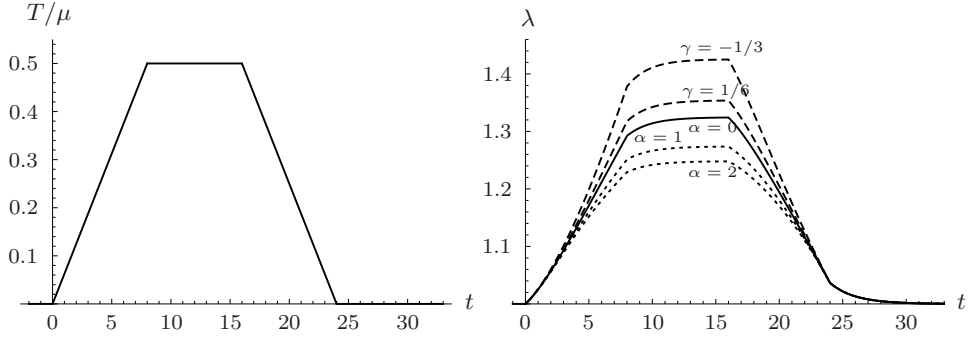}
			\caption{Plot of the dimensionless stress history $T/\mu$ (left) and the resultant  stretch $\lambda$ (right), from the Yeoh model predictions \eqref{stress_strain_qlv_yeoh} (dotted) and that from Mooney-Rivlin predictions \eqref{stress_strain_qlv_MR} (dashed). The continuous curve is  the neo-Hookean limit $\alpha=0$ (or $\gamma=1/2$).}
	\label{stress}
\end{figure}

An example is illustrated in Fig.\ \ref{stress}, with the imposed stress history shown on the left graph.
The one-term Prony series relaxation function \eqref{one_term_prony_series}, with constant values as used in the previous \S \ref{constants} is again employed. The resultant stretch is given on the right graph for the Yeoh strain energy function \eqref{stress_strain_qlv_yeoh} (dotted curves with $\alpha=1,2$), and for the Mooney-Rivlin strain energy function (dashed curves with $\alpha=1,2$). As before, the solid curve is the neo-Hookean prediction ($\alpha=0,\gamma=1/2$). As before, it can be seen that increasing $\alpha$ in the Yeoh model leads to material hardening, whilst decreasing $\gamma$ leads to softening of the Mooney-Rivlin material.

\subsection{Simple extension of a compressible bar} \label{simple_extension_compressible}

The purpose of this section is to repeat that analysis in \S \ref{elongation}\ref{simple_extension}
for a compressible material. We assume, as before, simple homogeneous extension of a bar, with zero tractions on its lateral faces, which is caused by an imposed stress or stretch in the $X_1$ direction, say. Then, the principal stretches may be given by
\begin{equation}
x_1(t)=\lambda_1(t) X_1,\quad x_2(t)=\lambda_2(t) X_2,\quad x_3(t)=\lambda_2(t) X_3,
\end{equation}
where symmetry between the $X_2$ and $X_3$ directions is clear, and so the deformation gradient tensor is
\begin{equation}\label{F_compr}
\mathbf{F}(t)=\textrm{diag}\left(\lambda_1(t),\lambda_2(t),\lambda_2(t)\right)
\end{equation}
and the left Cauchy Green tensor $\mathbf{B}$ becomes
\begin{equation}\label{F_B_compr}
\mathbf{B}(t)=\textrm{diag}\left(\lambda_1^2(t),\lambda^2_2(t),\lambda^2_2(t) \right).
\end{equation}
The principal invariants are therefore
\begin{align}
& I_1(t)= \lambda_1(t)^2+2 \lambda_2(t)^2, \qquad I_2(t)=2  \lambda_1(t)^2 \lambda_2(t)^2+ \lambda_2(t)^4, \qquad I_3=\lambda_1(t)^2  \lambda_2(t)^4. \label{invariants_simple_extens_compr}
\end{align}

The above can be substituted into the governing viscoelastic equation \eqref{cauchy_viscoelastic_stress_integrated},
with effective elastic stresses \eqref{piolaK_d_SEF}-\eqref{piolaK_h_SEF}, and simplified. The $X_1$ component of the stress yields, after simplification,
\begin{multline}
T(t)=\frac{\lambda_1(t) }{{\lambda_2}(t)^2} \Bigg[ \frac{4}{3}\int_0^t   \mathcal{D}'(t-s)  \left(1- \dfrac{\lambda_2^2(s)}{\lambda_1^2(s)}\right)  \left( W_1(s)+ W_2(s)  \lambda_2^2(s)\right)  \, \d s + \\  \dfrac{2}{3}\int_0^t  \mathcal{H}'(t-s)  \left(  W_1(s)  +2 \left( \dfrac{W_1(s)}{\lambda_1^2(s)}+2  W_2(s)  \right)\lambda_2^2(s) + \left(2  \dfrac{W_2(s)}{ \lambda_1^2(s)}+3  W_3(s) \right) \lambda_2^4(s)\right)   \, \d s \\ + 2 \left( W_1(t)   +2  W_2(t)  \lambda_2^2(t)+ W_3(t)  \lambda_2^4(t) \right)\Bigg], \label{first_compr_eq}
\end{multline}
and the normal stresses in the $X_2$ and $X_3$ equations are both
\begin{multline}
0=  \frac{1}{\lambda_1(t)}   \Bigg[\dfrac{2}{3}\int_0^t   \mathcal{D}'(t-s)  \left( 1-\dfrac{\lambda_1^2(s)}{\lambda_2^2(s)} \right)  \left( W_1(s)+ W_2(s)  \lambda_2^2(s)\right)  \, \d s \\+ \dfrac{2}{3}\int_0^t  \mathcal{H}'(t-s)  \bigg(  W_1(s) \dfrac{ \lambda_ 1^2(s)}{\lambda_2^2(s)}+ 2 \left( W_1(s) +2  W_2(s) \lambda_1^2(s) \right) + \\ \left(2  W_2(s)+3  W_3(s)  \lambda_1^2(s)\right) \lambda_2^2(s) \bigg)   \, \d s \\  + 2\left( W_1(t) + W_2(t)  \lambda_1^2(t)  +  \left(W_2(t) + W_3(t)\lambda_1^2(t)\right)  \lambda_2^2(t)\right)\Bigg], \label{second_compr_eq}
 \end{multline}
where as before $W_j$ is the derivative of $W$ with respect to the indicated invariant \eqref{W_derivs}. Clearly, the particular choice of the strain energy function affects the results and so, as before, a couple of examples are presented. Assuming the Horgan-Murphy strain energy function \cite{horgan-murphy08}, used for modelling a material  with a small compressibility, the SEF is
\begin{equation}\label{horgan-murphy}
W = \frac{\mu}{2}\left(\frac{1}{2}+\gamma\right)(I_1-3 I_3^{1/3})+\frac{\mu}{2}\left(\frac{1}{2}-\gamma\right)(I_2-3 I_3^{2/3})+\frac{\kappa}{2} (I_3^{1/2}-1)^2
\end{equation}
where $\gamma$ is an arbitrary constant, $\mu$ is the infinitesimal shear modulus and $\kappa$ is the infinitesimal bulk modulus. Then
\begin{equation} \nonumber
W_1=\dfrac{\mu}{2}\left(\frac{1}{2}+\gamma\right),\quad  W_2=\dfrac{\mu}{2}\left(\frac{1}{2}-\gamma\right),
\end{equation}
and
\begin{align}\nonumber
W_3&=-\dfrac{\mu}{2}\left(\frac{1}{2}+\gamma\right)I_3^{-2/3}-
\mu\left(\frac{1}{2}-\gamma\right)I_3^{-1/3}+\dfrac{\kappa}{2}\left(1-I_3^{-1/2}\right),
\end{align}
which are substituted in \eqref{first_compr_eq} and \eqref{second_compr_eq}. Setting $\gamma=1/2$, for ease of presentation, the stresses become
\begin{multline}
T(t)=\frac{1}{ \lambda_1(t)\lambda_2^2(t)}\Bigg[ \lambda_1^2(t) \Bigg(\mu+\frac{2 \mu }{3} \int_0^t  \mathcal{D}'(t-s) \left( 1- \dfrac{\lambda_2^2(s)}{\lambda_1^2(s)}\right)\ \d s + \\ \dfrac{1}{3} \int_0^t  \mathcal{H}'(t-s) \left(\mu \left(2\dfrac{ \lambda_2^2(s)}{\lambda_1^2(s)}+ 1 -3\dfrac{\lambda_2^{4/3}(s)}{\lambda_1^{4/3}(s)} \right) +3 \kappa \left(\lambda_2^4(s) - \dfrac{\lambda_2^2(s)}{\lambda_1(s)}\right)\right)\ \d s + \kappa \lambda_2^4(t) \Bigg) \\
- \lambda_1^{2/3}(t) \lambda_2^{4/3}(t) \left(\mu +\kappa \lambda_1^{1/3}(t) \lambda_2^{2/3}(t) \right) \Bigg]   \label{first_compr_eq_HM}
\end{multline}
and
\begin{multline}
0=\frac{1}{\lambda_1(t) \lambda_2^2(t)}\Bigg[ \lambda_2(t)^2\Bigg(\mu+\frac{\mu}{3}\int_0^t \mathcal{D}'(t-s)  \left( 1-\dfrac{\lambda_1(s)^2}{\lambda_2^2(s)} \right)\ \d s \\ + \dfrac{1}{3} \int_0^t \mathcal{H}'(t-s)    \left(\mu \left(2 +\dfrac{\lambda_1^2(s)}{\lambda_2^2(s)} -3 \dfrac{\lambda_1^{2/3}(s)}{\lambda_2^{2/3}(s)} \right)+3 \kappa \lambda_1(s)\left(\lambda_1(s)\lambda_2^2(s)-1\right)\right) \ \d s\Bigg) + \\   \kappa \lambda_1^2(t)    \lambda_2^4(t) - \lambda_1^{2/3}(t)  \lambda_2^{4/3}(t) \left(\mu +\kappa \lambda_1^{1/3}(t)  \lambda_2^{2/3}(t) \right) \Bigg], \label{second_compr_eq_HM}
\end{multline}
respectively. By contrast, for the Gent model (see for example \cite{macdonald2007practical}),
\begin{equation}
W=\dfrac{\mu}{2} J_m \log\left(1-\dfrac{I_1-3}{J_m}\right)^{-1}+\dfrac{\kappa}{2}\left(\dfrac{I_3-1}{2}- \dfrac{1}{2}\log I_3\right)^4,   \nonumber
\end{equation}
where $J_m$  is a constant limiting value for $I_1-3$, taking into account limiting polymeric chain extensibility. Hence
\begin{equation}
W_1=\frac{\mu }{2} \left(1-\frac{I_1-3}{ J_m}\right)^{-1},\quad W_2=0,\quad W_3=\dfrac{ \kappa}{8}\left(1-\dfrac{1}{I_3}\right)  \left(I_3-1- \log I_3\right)^3,
\end{equation}
which can then substituted into \eqref{first_compr_eq} and \eqref{second_compr_eq} to yield the equivalent results for the stresses. These are omitted for brevity.

\section{Concluding remarks}\label{conclusion}

This article has focused on reappraising Fung's method for quasilinear viscoelasticity. It has been shown that some of the negative features commonly associated with the approach are merely a consequence of the way it has been applied elsewhere. The approach outlined herein was shown in \S \ref{elongation} to yield \textit{sensible} results, and offers a straightforward approach to solving a wide range of models. The present method exhibits similarities with Simo's approach to nonlinear viscoelasticity \cite{simo1987}, although the latter assumed a scalar relaxation function acting on the stress. Therefore, it is noted that Simo's method would not reduce in the linear limit to the usual relation \eqref{linear_viscoelasticity_integrated} if the shear and bulk moduli have different temporal behaviour. (Note that ABAQUS version 6.12 employs the same relaxation constants in both the shear and bulk parts of the field.) Herein the relaxation function is a tensor, which for isotropic materials reduces to two distinct scalar relaxation functions, one acting on the compressive part and the other on the shear component of the stress. This will therefore allow consistency with the linear theory. 

It was shown that for an imposed stretch, or torsion, it is simple matter to solve the QLV equation directly. For imposed stress, the stretch has to be deduced by solving (inverting) the integral equation, and this is achieved here using a discretised scheme accurate to an order of the cube of time-step size. Muliana et al.\ have recently offered a quasilinear viscoelastic model \cite{muliana-rajagopal-wineman13} where the strain is expressed as a function of the stress, which may be viewed as a dual model to \eqref{quasi_linear_constitutive_law}. However, it is not clear as yet whether their approach offers as effective an approach at modelling viscoelasticity as Fung's scheme.

The authors are currently applying the present approach to a range of problems in viscoelasticity. The numerical procedure described in Appendix A can be employed simply for any deformation that is incompressible and homogenous, for example simple shear. It is presently being extended so that it can solve for compressible materials undergoing simple extension or shear. The QLV method is also adaptable to inhomogeneous problems, for example large amplitude radial deformations in 2D and 3D, and simple torsion. It is anticipated that the present approach will be more suited to satisfying the equations of equilibrium than, say, the more heuristic QLV relation employed in ABAQUS [v.6.12]. Finally, the authors are using the model for several biomechanics problems, to derive a perturbation theory for the viscoelastic evolution of small deformations on top of large ones, and to derive effective properties of ligaments and other tissues.   

\section*{Acknowledgment}

The authors are grateful to the Engineering and Physical Research Council for the award (grant number EP/H050779/1) of a postdoctoral research assistantship for De Pascalis.

\appendix
\section{A numerical high-order solution procedure for Volterra integral equations} \label{numeric}
The purpose of this Appendix is to offer a simple and rapid numerical scheme for solving Volterra integral equations of the type obtained by the QLV approach (see e.g.\ \eqref{stress_strain_qlv_yeoh}).  We assume they   have the separable form
\begin{equation}\label{general_num_eq2}
T(t)=g(\lambda(t))+\sum_{j=1}^{N} f_j(\lambda(t))\int_0^t \mathcal{D}'(t-s) h_j(\lambda(s))\ \d s,
\end{equation}
where $T(t), \mathcal{D}(t), g(X), f_j(X)$ and $h_j(X)$ are all known functions of their respective arguments. Clearly, all the equations offered in \S \ref{elongation}\ref{simple_extension} lie within this class but for more general problems, with inhomogeneous deformations, the analysis described below must be amended.

Let us now discretise the system, and evaluate the equation at each time $t_n=n\delta t$, with $n=0,1,2,3,\ldots,n_{\textrm{max}}$, where $\delta t$ is a \textit{small} time step. On writing the stretch at each time step as
\begin{align}
&\lambda_n=\lambda(t_n)=\lambda(n\delta t),
\end{align}
a Taylor series expansion can be employed to yield
\begin{multline} \label{g_expansion}
g(\lambda_n)=g(\lambda_{n-1})+g'(\lambda_{n-1}) (\lambda_n-\lambda_{n-1}) + \dfrac{g''(\lambda_{n-1})}{2!}(\lambda_n-\lambda_{n-1})^2+\\  \dfrac{g'''(\lambda_{n-1})}{3!}(\lambda_n-\lambda_{n-1})^3+O((\lambda_n-\lambda_{n-1})^4),
\end{multline}
where $'$ denotes differentiation with respect to the argument, and
\begin{equation}\label{lam_expans}
\lambda_n=\lambda(n\delta t)=\lambda_{n-1}+ \lambda' _{n-1} \delta t  +\dfrac{\lambda''_{n-1} }{2!}\delta t^2 + \dfrac{\lambda'''_{n-1} }{3!} \delta t^3+O(\delta t^4).
\end{equation}
Note that all functions are assumed sufficiently smooth between adjacent time steps so that the relations above and below hold. From \eqref{lam_expans}, it is clear that
\begin{align}
\left(\lambda_{n}-\lambda_{n-1}\right)&=    \lambda' _{n-1}\delta t+\dfrac{\lambda''_{n-1}}{2!}\delta t^2 + \dfrac{\lambda'''_{n-1} }{3!}\delta t^3 +O(\delta t^4),\notag \\
 \left(\lambda_{n}-\lambda_{n-1}\right)^2&= (\lambda'_{n-1})^2 \delta t^2 +  \lambda'_{n-1}\lambda''_{n-1}  \delta t^3 +O(\delta t^4),  \label{lamdiff} \\
 \left(\lambda_{n}-\lambda_{n-1}\right)^3&=    (\lambda'_{n-1})^3\delta t^3+O(\delta t^4).\notag
\end{align}
The integral in \eqref{general_num_eq2} can also be expanded as
\begin{align}
\int_0^{t} \mathcal{D}'(t-s) h_j(\lambda(s))\ \d s&=\int_0^{n\delta t} \mathcal{D}'(n \delta t-s) h_j(\lambda(s))\ \d s     \notag \\ &= \sum_{m=0}^{n-1} \int_{m\delta t}^{(m+1)\delta t} \mathcal{D}'(n \delta t-s)  h_j(\lambda(s))\ \d s \notag \\ &=  \sum_{m=0}^{n-1} \int_{0}^{\delta t} \mathcal{D}'((n-m) \delta t-u)  h_j(\lambda(m\delta t+u))\ \d u
\end{align}
where in the latter the substitution $s=m\delta t+u$ is used. The integrand functions can be further expanded as
\begin{align}
\mathcal{D}'\left((n-m)\delta t-u\right) =  \mathcal{D}'\left((n-m)\delta t\right)- \mathcal{D}''\left((n-m)\delta t\right) u+\dfrac{\mathcal{D}'''\left((n-m)\delta t\right) }{2!}  u^2  +O(u^3)
\end{align}
and
\begin{align}
h_j\left(\lambda(m\delta t+u)\right)&= h_j (\lambda_m)+ \dfrac{\d h_j(\lambda_m)}{\d u}  u+ \dfrac{1}{2!} \dfrac{\d^2 h_j(\lambda_m)}{\d u^2}u^2 +O(u^3) \notag \\
&=h_j (\lambda_m)+  \lambda_m'     h_j'  (\lambda_m)u+ \dfrac{1}{2!}  \left(\lambda''_mh'_j(\lambda_m)+(\lambda'_m)^2 h''_j(\lambda_m)\right)u^2+O(u^3),
\end{align}
and therefore, after some simple algebra and integration, it is found that
\begin{multline}
\int_0^{\delta t}  \mathcal{D}'\left((n-m)\delta t-u\right)  h_j\left(\lambda(m\delta t +u)\right)\d u=   \mathcal{D}'\left((n-m)\delta t \right)  h_j\left(\lambda_m\right) \delta t+\\ \left( \mathcal{D}'\left((n-m)\delta t\right) \lambda'_m h'_j(\lambda_m)- \mathcal{D}''\left((n-m)\delta t\right)h_j(\lambda_m) \right) \dfrac{\delta t^2}{2} \\ +  \Big( \mathcal{D}'''\left((n-m)\delta t\right)h_j(\lambda_m)-2 \mathcal{D}'' \left((n-m)\delta t\right)\lambda'_m h_j' (\lambda_m)+ \\ \mathcal{D}'\left((n-m)\delta t\right) \left(\lambda''_m h'_j(\lambda_m)+(\lambda'_m)^2 h''_j(\lambda_m)\right)\Big) \dfrac{\delta t^3 }{6} +O(\delta t^4).\\
\end{multline}
Using now the expansion given in \eqref{g_expansion}, with the help of the relation \eqref{lamdiff}, it is possible to rewrite the expansion both for $g$ and for $f_j$ as
\begin{multline}
g(\lambda_n)=g(\lambda_{n-1})+ \lambda'_{n-1} g'(\lambda_{n-1}) \delta t+ \left(\lambda''_{n-1}g'(\lambda_{n-1})+(\lambda'_{n-1})^2 g''  (\lambda_{n-1})\right) \dfrac{\delta t^2}{2}+  \\ \left(\lambda'''_{n-1}g'(\lambda_{n-1})+3\lambda'_{n-1}\lambda''_{n-1}g''(\lambda_{n-1})+(\lambda'_{n-1})^3g'''(\lambda_{n-1})\right) \dfrac{\delta t^3}{6} +O(\delta t^4),\notag
\end{multline}
and
\begin{multline}
 f_j(\lambda_n)=f_j(\lambda_{n-1})+\lambda'_{n-1} f'_j(\lambda_{n-1}) \delta t +\left(\lambda''_{n-1}f'_j(\lambda_{n-1})+(\lambda'_{n-1})^2 f''_j(\lambda_{n-1})\right)\dfrac{\delta t^2}{2} +\\  \left(\lambda'''_{n-1}f_j'(\lambda_{n-1})+3\lambda'_{n-1}\lambda''_{n-1}f_j''(\lambda_{n-1})+(\lambda'_{n-1})^3f_j'''(\lambda_{n-1})\right) \dfrac{\delta t^3}{6}+O(\delta t^4). \notag
\end{multline}
Finally, gathering all terms, the whole equation \eqref{general_num_eq2} can be written as the following expansion
\begin{multline} \label{full_expans}
T(n\delta t)= g(\lambda_{n-1})+ \lambda'_{n-1} g'(\lambda_{n-1}) \delta t+   \left(\lambda''_{n-1}g'(\lambda_{n-1})+(\lambda'_{n-1})^2 g''(\lambda_{n-1})\right) \dfrac{\delta t^2}{2}+ \\ \left(\lambda'''_{n-1}g'(\lambda_{n-1})+3\lambda'_{n-1}\lambda''_{n-1}g''(\lambda_{n-1})+(\lambda'_{n-1})^3g'''(\lambda_{n-1})\right) \dfrac{\delta t^3}{6} +\\  A_{n-1} \delta t +  B_{n-1} \dfrac{\delta t^2}{2}+ C_{n-1} \dfrac{\delta t^3}{6} +O(\delta t^4),
\end{multline}
where $A_{n-1}, B_{n-1} , C_{n-1}$ are the following expressions
\begin{equation}
A_{n-1} = \sum_{j=1}^N\sum_{m=0}^{n-1} \Big[ f_{j}(\lambda_{n-1}) \mathcal{D}'\left((n-m)\delta t\right)h_j(\lambda_m)\Big], \\
\end{equation}
\begin{multline}
B_{n-1} = \sum_{j=1}^N\sum_{m=0}^{n-1} \Big[ 2 \lambda'_{n-1} f'_j(\lambda_{n-1}) \mathcal{D}'\left((n-m)\delta t\right) h_j(\lambda_m) \\ + f_j(\lambda_{n-1}) \Big(\mathcal{D}'\left((n-m)\delta t\right) \lambda'_mh'_j(\lambda_m)-\mathcal{D}''\left((n-m)\delta t\right)h_j(\lambda_m)\Big)\Big],
\end{multline}
\begin{multline}
C_{n-1} =\sum_{j=1}^N\sum_{m=0}^{n-1} \Big[  f_j(\lambda_{n-1}) \Big(\mathcal{D}'''\left((n-m)\delta t\right) h_j(\lambda_m)-2 G''\left((n-m)\delta t\right) \lambda'_m h'_j (\lambda_m)+\\ \mathcal{D}'\left((n-m)\delta t\right) \left(\lambda''_m h'_j(\lambda_m)+(\lambda'_m)^2h''_j (\lambda_m)\right)\Big) \\+ 3 \lambda'_{n-1} f'_j(\lambda_{n-1}) \Big(\mathcal{D}'\left((n-m)\delta t\right) \lambda'_m h'_j(\lambda_m)-\mathcal{D}''\left((n-m)\delta t\right) h_j(\lambda_m)\Big) \\+ 3 \mathcal{D}'\left((n-m)\delta t \right)  h_j\left(\lambda_m\right) \Big(\lambda''_{n-1}f'_j(\lambda_{n-1})+(\lambda'_{n-1})^2f_j''(\lambda_{n-1})\Big) \Big].
\end{multline}
This equation must hold for each $n$, and at each level of approximation. Hence, keeping terms up to $O(\delta t)$ in \eqref{full_expans}, and rearranging, gives
\begin{equation}\label{lambda_dash}
\lambda'_{n-1}=\left(T(n\delta t)- g(\lambda_{n-1}) - A_{n-1} \delta t \right)/\left(\delta t g'(\lambda_{n-1})\right),
\end{equation}
and then equating terms at higher order ($O(\delta t^2)$ and $O(\delta t^3)$) to zero yields, respectively,
\begin{align} \label{lambda_dash2}
\lambda''_{n-1}&=-\bigg((\lambda'_{n-1})^2 g''(\lambda_{n-1})+B_{n-1}  \bigg)/g'(\lambda_{n-1}),\\ \label{lambda_dash3}
\lambda'''_{n-1}&=-\bigg(3  \lambda'_{n-1}\lambda''_{n-1}g''(\lambda_{n-1})+(\lambda'_{n-1})^3 g'''(\lambda_{n-1}) +  C_{n-1}  \bigg)/g'(\lambda_{n-1}).
\end{align}

Now, the procedure for solving the discretised equations \eqref{full_expans}, for specified stress, is clear. We start at $t=0$ (or $n=0$) and so from \eqref{general_num_eq2}:
\begin{equation}\label{t_0}
T(0)=g(\lambda(0))
\end{equation}
gives
\begin{equation}\label{t_0_inverse}
\lambda(0)=\lambda_0= g^{-1} (T(0)).
\end{equation}
As there is no stretch prior to an applied stress, an initial zero stress, $T(0)=0$, would mean that $\lambda_0=1$. The derivatives of $\lambda_0$
are derived from \eqref{lambda_dash}-\eqref{lambda_dash3} (setting $n=1$ in these equations) and hence $\lambda_1$ is found from \eqref{lam_expans}.
Then $n$ is incremented, the values of $\lambda', \lambda'', \lambda'''$ determined, and these are again used to determine the stretch at the next time step. The process is repeated at each following step. Note that size of the sums in $A_{n-1}, B_{n-1}, C_{n-1}$ grow with $n$, and so for large times, or using small time steps, evaluation of the system can become rather slow. However, this difficulty can be alleviated, with the present system, by obtaining a relationship between $A_{n}$ and its previous increment $A_{n-1}$, and similarly for $B_{n}, C_{n}$. The precise forms of these relationships are determined from the given $f_j(\lambda)$ and the relaxation function $\mathcal{D}(t)$. In practice, the solution procedure works well, with an error of  $O(\delta t^4)$, i.e.\ for a time step of $\delta t=0.01$ then the error is $O(10^{-8})$.


\bibliographystyle{stylericcardo}
\bibliography{bibliography}

\end{document}